\documentstyle[11pt,twoside,jltp]{article}

\title{Induced Gravity in Superfluid  $^3$He}

\author{G.E. Volovik\address{Helsinki University of Technology,  Low
Temperature
Laboratory, P.O.Box 2200, FIN-02015 HUT, Finland} \address{permanent address:
Landau Institute for Theoretical Physics, Moscow, Russia}}

\runninghead{G.E. Volovik}{Induced Gravity in Superfluid  $^3$He}

\begin{document}

\begin{abstract}
The gapless fermionic excitations in superfluid
$^3$He-A have the "relativistic" spectrum close to the gap nodes. This
allowed us
to model the modern cosmological scenaria of baryogenesis  and
magnetogenesis. The same massless fermions induce another low-energy
property of the quantum vacuum -- the gravitation. The effective metric of the
space, in which the free quasiparticles move along geodesics, is not generally
flat. Different order parameter textures correspond to curved effective
space and produce many different exotic metrics, which are theoretically
discussed in quantum gravity and cosmology. This includes the condensed matter
analog of the black hole and event horizon, which can be realized in the moving
soliton. This will allow us to simulate and thus experimentally investigate
such quantum phenomena as the Hawking radiation from the horizon, the
Bekenstein
entropy of the black hole, and the structure of the quantum vacuum behind the
horizon. One can also simulate the conical singularities produced by cosmic
strings and monopoles; inflation; temperature dependence of the cosmological
and Newton constants, etc.

\end{abstract}

\maketitle


\section{Introduction}

 The quantum physical vacuum -- the former ether
-- appears to be a complicated substance. Our present experimental physics is
able to investigate only its long-wave-length properties. The highest energy
of the elementary particles, i.e. of the elementary excitations of the physical
vacuum, which is achieved today, is much smaller than the characteristic
Planck value $E_P=\sqrt{\hbar c^5/G}$, where $c$ is a speed of light and
$G$ the
Newton gravitational constant. Their wave lengths are correspondingly  much
larger than the Planck scale $r_P =\hbar c/E_P  \sim 10^{-33}$cm, which
characterizes the ``microscopic'' structure of the vacuum -- the ``Planck
solid'' \cite{Jegerlehner} or ``Planck condensed matter''. Thus we are
essentially limited in probing the microscopics of vacuum.

On the other hand we know that the low-energy
properties of different condensed matter substances (magnets, superfluids,
crystals, superconductors, etc.)  are robust, i.e. do not depend much on the
details of the microscopic structure of these substances. The main role is
played by symmetry and topology of condensed matter: they determine the soft
(low-energy) hydrodynamic variables, the effective Lagrangian describing
the low-energy dynamics, and topological defects. The microscopic details
provide us only with the ``fundamental constants'', which enter the effective
Lagrangian, such  as speed of sound, superfluid density, modulus of elasticity,
magnetic susceptibility, etc.

According to this analogy,  such properties of our world, as gravitation, gauge
fields, elementary chiral fermions, etc., arise as a low-energy soft modes
of the
``Planck condensed matter''. At high energy (of the Planck scale) these
soft modes disappear: actually they merge with the continuum of the high-energy
degrees of freedom of the ``Planck condensed matter'' and thus cannot be
separated anymore.

The main advantage of the condensed matter analogy is that in principle we know
the condensed matter structure at any relevant scale, including the interatomic
distance,  which corresponds to the Planck scale. Thus the condensed matter can
serve as a guide on the way from our present low-energy
corner to the Planckian and trans-Planckian physics. In this sense the
superfluid phases of
$^3$He, especially $^3$He-A, are of most importance: the low-energy degrees of
freedom in $^3$He-A do really consist of chiral fermions, gauge fields and
gravity. That is why, though there is no one-to-one correspondence between the
$^3$He-A and the ``Planck condensed matter'', many phenomena of quantum vacuum
can be simulated in $^3$He-A.

In a previous QFS-97 talk \cite{VolovikQFS97} we discussed the possibility of
using the superfluid phases of $^3$He as a model for the investigation of a
property of the quantum vacuum known as chiral anomaly, which leads to the
nonconservaton of the baryonic charge and thus to the possibility of generation
of the baryonic asymmetry of our present Universe. The $^3$He
analogue of the chiral anomaly was verified in experiments
with quantized vortices.\cite{BevanNature} After QFS-97 it became clear
that yet
another  effect observed in $^3$He-A -- collapse of the normal-to-superfluid
counterflow with formation of textures
\cite{CounterflowCollapse} -- is related to the phenomenon of the chiral
anomaly
\cite{KVV,VolovikLammi}. Moreover it appeared that this collapse completely
reproduces one of the scenaria of  the generation of  galactic magnetic fields
in the early Universe.\cite{JoyceShaposhnikov,GiovanniniShaposhnikov} Another
direction which was developed in the period between QFS-97 and QFS-98  is
related to the gravitation: how the gravity can be viewed from the
$^3$He-A physics and how different problems of gravity can be experimentally
simulated in $^3$He-A.

\section{Gap Nodes and Gravity}

The effective gravity, as a low-frequency phenomenon, arises in
many condensed matter systems. In crystals, the effective curved space, in
which
``elementary particles'' -- phonons and Bloch electrons -- are propagating,
is induced by the distributed topological defects, dislocations and
disclinations, (see e.g.
\cite{Katanaev,Baush,Dzyal}). In the normal (or superfluid) liquids, the
effective Lorentzian space for propagating sound waves (phonons) is
generated by
the hydrodynamic (super) flow of the liquid.\cite{UnruhSonic,Visser1997}
However it appears that the superfluid $^3$He-A provides us with the most
adequate analogy of the relativistic theory of the effective gravity, first
introduced by Sakharov \cite{Sakharov}.

The most important property of the
superfluid $^3$He-A, as compared to other quantum fluids and solids, is
that its elementary fermions -- Bogoliubov-Nambu quasiparticles -- are gapless.
Their energy spectrum $E({\bf p})$ contains the point nodes, where the
energy is
zero. Close to the gap nodes, i.e. at low energy, these quasiparticles have a
well defined chirality: the quasiparticles are either left-handed or
right-handed. This effectively reproduces our low-energy world (however above
the electroweak energy scale), where all the fermionic elementary particles
(electrons, neutrinos, quarks) are chiral. The point gap nodes are
topologically
stable and thus do not disappear under the deformation of the system, the gap
nodes with oppposite topological charges can however annihilate each other
\cite{VolovikLammi}. If the total topological charge of the nodes is nonzero
some of the gapless (massless) fermions will survive under any perturbation of
the system. This algebraic conservation of massless fermions induced by
the momentum space topology can be a good reason for the zero mass of neutrino:
If we live in the vacuum  with nonzero total topological charge,  we should not
spend time and money to look for the neutrino mass.  Recent
Kamiokande experiments have shown evidence of neutrino oscillations and
thus the
possibility of the neutrino mass.\cite{Kamiokande} If this is true, this would
mean that the total topological charge of the nodes in the ``Planck condensed
matter''  is zero. We shall see.

Another interesting consequence of the existence of the gap
node is that it immediately introduces the effective gravitational field.
Let the
node be situated at point
${\bf
p}^{(0)}$ in momentum space, i.e. $E({\bf p}^{(0)})=0$. If the space
parity is not violated, then in the vicinity of the node the function $E^2({\bf
p})$ takes a quadratic form for deviations from ${\bf p}^{(0)}$:
\begin{equation}
E^2({\bf p})=g^{ik}(p_i -p^{(0)}_i) (p_k - p^{(0)}_k)~.
\label{E^2form}
\end{equation}
The quantity $g^{ik}$ plays the part of the effective
metric tensor of the gravitational field, while ${\bf A}={\bf p}^{(0)}$
corresponds to the vector potential of effective electromagnetic field.
The other dynamical components of these fields appear, if
the space parity is violated,  for example if the superflow velocity ${\bf
v}_s\neq 0$. In this case the quasiparticle energy is Doppler shifted,
$E({\bf p})
\rightarrow  E({\bf p}) +{\bf p}\cdot {\bf v}_s = E({\bf
p}) +{\bf p}^{(0)}\cdot {\bf v}_s + ({\bf p}-{\bf
p}^{(0)})\cdot {\bf v}_s$. This leads to appearance of the scalar vector
potential $A_0={\bf p}^{(0)}\cdot {\bf v}_s$ and the mixed component of the
metric tensor $g^{i0}=v_s^i$. With these identifications, Eq.(\ref{E^2form})
becomes fully ``relativistic'':
\begin{equation}
 g^{\mu\nu}(p_\mu -eA_\mu) (p_\nu -eA_\nu)=0~.
\label{rel}
\end{equation}
Here the ``electric charge'', $e=\pm$, reflects the fact that in $^3$He-A each
gap node has a sparring partner with an opposite momentum, $-{\bf p}^{(0)}$:
quasiparticles near the opposite nodes have opposite charges (and also
the opposite chiralities).

From the topological stability of the gap node in fermionic spectrum, it
follows
that the small  deformations of the quantum vacuum do not destroy the gap node,
but they lead to the variation of the fields $g^{\mu\nu}$ and $A_\mu$. This
means
that the ``gravitational metric'' and the ``gauge fields''  are dynamical
collective modes of the quantum vacuum in  fermionic condensed matter which
has the point gap nodes. Moreover, in the vicinity of the gap nodes the
fermions are chiral and are described by the Weyl equations. Thus in the
low-energy corner the $^3$He-A has all the ingredients of the quantum field
theory of our relativistic vacuum: chiral fermions, gauge fields and
gravitation. This suggests that  the ``Planck condensed matter'' belongs to the
same universality class as $^3$He-A  (though not all the components
of the gauge and graviational fields are independent  in $^3$He-A, see
\cite{VolovikLammi}). Within this picture, the internal
symmetries, such as $SU(2)$ and
$SU(3)$, are  consequences of the given number of gap nodes and the symmetry
relations between them: In $^3$He-A the $SU(2)$ gauge field naturally arises in
this manner (see \cite{Exotic}). The main difference from the ideologically
similar relativistic theories of induced gravity (such as in Ref.
\cite{FrolovFursaev}, where the low-energy gravity is induced by massive
relativistic  quantum fields) is that in our case the gravity appears in the
low-energy corner simultaneously with ``relativistic'' fermions and bosons.

\section{Charge of Vacuum and Cosmological Term}

An equilibrium homogeneous ground state of condensed matter has zero charge
density, if charges interact via long range forces. For example,
electroneutrality is the necessary property of bulk metals and
superconductors; otherwise the vacuum energy of the system diverges faster
than its volume. Another example is the algebraic density of quantized
vortices in superfluids, $<\nabla \times {\bf v}_s>$: due to logarithmic
interaction of vortices it must be zero in  equilibrium homogeneous
superfluids (in the absence of external rotation). The same argument can be
applied to the ``Planck condensed matter'', and it seems to work. The
density of
the electric charge of the Dirac sea is zero due to exact cancellation of
electric charges of electrons,
$q_e$, and quarks,
$q_u$ and $q_d$, in the fermionic vacuum:
$Q_{\rm vac}=\sum_{E<0}(q_e+3q_u+3q_d)=0$. In Einstein theory the
energy-momentum tensor of the vacuum should be a source of the long range
gravitational forces. The immediate consequence is that for the
equilibrium homogeneous state of the vacuum one has $\partial S_{\rm
vac}/\partial g_{\mu\nu}= \sqrt{-g}T^{\mu\nu}_{\rm vac}=0$ (this
equilibrium condition  contains partial derivative, instead of functional
derivative).  This leads to nullification of the cosmological term in
Einstein equations for vacuum in equilibrium, which agrees with the experiment.
But it is in very serious disagreement with  theoretical estimations of
fermionic
or bosonic zero-point vacuum energy, which is 120 orders of magnitude
higher than
the experimental upper limit. What is wrong with theory?

The situation is probably similar to what we have in condensed matter if we
want to estimate the ground-state energy from the zero-point energy of phonons:
$E_{\rm zp}=(1/2)\sum_{\bf k} \hbar\omega ({\bf k})$. This never gives the
correct estimation of the vacuum energy; moreover sometimes it has even the
wrong
sign. This occurs because the phonon modes are soft variables and are
determined only in the low-energy limit, whereas the vacuum energy of the
solid is determined by the quantum many-body physics, where essentially the
high-energy degrees of freedom are involved. These
high-energy degrees are always adjusted to provide the electroneutrality,
irrespective of the low-energy physics. This suggests two consequences:

(1) The $^3$He analogy suggests  that a zero value of the cosmological term in
the equilibrium vacuum is dictated by the Planckian or trans-Planckian
degrees of
freedom:  $\partial S_{\rm vac}/\partial g_{\mu\nu}=0$ is the thermodynamic
equilibrium condition for the ``Planck condensed matter''. Thus the
equilibrium homogeneous vacuum does not gravitate. Deviations of the vacuum
from
its equilibrium can gravitate
\cite{GravityIn3He-B}. For example, in the presence of  matter, a small
cosmological term of the order of the energy density of the matter is possible.

(2) The quantization of the low-energy degrees of freedom, such as
the gravitational field, works only in the low-energy limit, but cannot be
applied to higher energies. Moreover, even in the low-energy limit the
quantization leads to the double counting: The total energy of the solid (and
the same can be applied to the ``Planck solid'') is obtained by the solution of
the quantum many-body problem, which automatically takes into account all the
degrees of freedom.  If one now separates the soft modes and adds the
zero-point
energy of these modes, this will be the double counting. The gravity is a
low-frequency, and actually classical result of quantization of high-energy
degrees of freedom of the ``Planck condensed matter'', so one should not
quantize the gravity again.  The quantization of gravity is similar to the
attempt to derive the microscopic structure of the solid, using only the
spectrum of the low-energy phonons. The $^3$He-A analogy also suggests that all
the degrees of freedom, bosonic and fermionic, can come from the initial (bare)
fermionic degrees of freedom.

\section{Gravitational Constant in $^3$He-A.}

$^3$He-A is not a perfect system for the modelling of all the aspects of
gravity.
The effective Lagrangian for $g_{\mu\nu}$, which is obtained after integration
over the fermionic degrees of freedom, contains many noncovariant terms (see
discussion in \cite{VolovikLammi,GravityOfMonopole}). However some terms
correspond to the Einstein action, and this allows us to estimate the
gravitational constant $G$ in the effective theory. Consider an analog of
the graviton field in $^3$He-A. In the
relativistic gravitation the energy density of the graviton propagating along
$z$ direction is written in terms of the perturbation of the metric,
$g_{\mu\nu}=  g^{(0)}_{\mu\nu}+h_{\mu\nu}$:
\begin{equation}
 {\cal T}^0_0=   {1   \over
16 \pi G} \left[(\partial_z h_{xy})^2 + {1\over 4}((\partial_z
(h_{xx} -h_{yy}))^2 \right]  ~~.
\label{Graviton EnergyDensity}
\end{equation}
In $^3$He-A such a perturbation corresponds to the so-called clapping mode
with spin 2 \cite{Exotic,GravityOfMonopole}. The clapping mode and the graviton
have the same structure of the energy density; thus comparing coefficients in
expressions for ${\cal T}^0_0$ one obtains the temperature
dependent effective gravitational constant:\cite{GravityOfMonopole}
\begin{equation}
G(T)=  {12 \pi \over  K(T)\Delta^2(T)}~.
\label{GravitationConstant}
\end{equation}
Here $\Delta$ is the gap amplitude, which plays the part of the Planck
energy; $K(T)$ is a dimensionless function of $T$, which
close to the superfluid transition temperature $T_c$ is
\begin{equation}
 K(T) = 1-{T^2\over T_c^2} ~~,~~T\rightarrow T_c \sim \Delta (0)~.
\label{K(Tc)}
\end{equation}
The temperature dependence of $G$ corresponds to screening of gravity by the
fermionic vacuum. The dependence coming from the factor $K(T)$
is the traditional one: Since the effective action for gravity is obtained from
the integration over the fermionic (or bosonic) degrees of freedom
\cite{Sakharov}, it is influenced by the thermal distribution of fermions.
What is new here is that $G$ is also influenced by the temperature
dependence of
the ``Planck'' energy cut-off $\Delta(T)$, which is determined by details
of the
trans-Planckian physics. In $^3$He-A, $\Delta^2(T)\sim
\Delta^2(0) (1-T^2/T_c^2)$. This provides an illustration of how the
corrections $T^2/E_P^2$, caused by the Planckian physics, come into play even
at low $T$ \cite{Jegerlehner}.  So, we can hope, that even the present
``low-energy'' physics contains the measurable corrections coming from the
Planck
physics.

\section{Event Horizon}

At the classical level the existence of an event horizon leads to the
divergency
of the density of the particle states at the horizon (see e.g.
\cite{Padmanabhan}
and references therein). At the quantum level this is believed to give rise to
the Bekenstein entropy of the black hole,  and the
Hawking radiation from the black hole \cite{Bekenstein,Hawking}. There are many
problems related to these issues: Are the entropy and the Hawking radiation
physical?   Does the entropy come from degrees of freedom (i) outside
the horizon, (ii) on the horizon, (iii) inside the horizon? Since the particle
momentum grows up to the Planck scale at the horizon,  trans-Planckian physics
is inevitably involved. Also it is not completely clear whether the quantum
vacuum is stable in the region behind horizon.

These problems can be treated in condensed matter of the $^3$He-A type,
where in
the ``cis-Planckian'' scale the quasiparticles -- excitations of the quantum
vacuum -- are ``relativistic'' and obey the dynamical equations, determined by
the effective Lorentzian metric in Eq.(\ref{rel}). Existence  of
the quantum vacuum, which can respond to the change of the
fermionic spectrum in the presence of horizon, is instrumental and represents a
big advantage  of $^3$He-A compared with
conventional liquids and other dissipative systems
\cite{UnruhSonic,Visser1997,Rosu}, where the event horizon can also arise.

\subsection{Horizon within moving domain wall.}

In $^3$He-A the analog of event horizon appears in the moving topological
soliton \cite{JacobsonVolovik}. Here we consider soliton moving in a thin
film of
$^3$He-A. In the parallel-plane geometry the unit vector
${\hat{\bf l}}$, which determines the direction of the gap node in momentum
space, is fixed by the boundary conditions -- it should be normal to the film:
${\hat{\bf l}}=\pm{\hat{\bf z}}$ if the film is in the plane $x,y$ plane.
Due to the double degeneracy, a topologically stable domain wall
can exist which separates two half-spaces with opposite directions of
${\hat{\bf l}}$. The classical spectrum of the low-energy
fermionic quasiparticles in the presence of the domain wall is:
\begin{equation}
E^2({\bf p})=(c^z)^2 (p_z\mp p_F)^2+(c^x)^2 p_x^2 + (c^y)^2 p_y^2 ~~.
\label{E}
\end{equation}
In the simplest soliton the ``speeds of light''  propagating along $x,y,z$ are
\begin{equation}
c^z=v_F~,~c^y= c_\perp ~,~c^x(x)= c_\perp  \tanh {x\over d}  ~~,~
c_\perp={\Delta(T) \over p_F} ~,
\label{3SpeedsOfLight}
\end{equation}
where $p_F$ and $v_F$ are the Fermi momentum and Fermi velocity; the thickness
$d$ of the domain wall is on order of superfluid coherence length,
$d\sim \xi \sim v_F/\Delta(T)$.
In this domain wall the ``speed of light'' propagating across the
wall, $c^x(x)$, changes sign, while the other two remain constant.

If such a domain wall is moving with velocity $v$ along $x$,
the Doppler shifted fermionic spectrum gives the
following time-independent metric in the frame of the moving wall:
\begin{equation}
g^{zz}= v_F^2 ~,~g^{yy}= c_\perp^2~,~g^{xx}= (c^x(x))^2
-v^2~,~g^{00}=-1,~g^{0x}=v~.
\label{Metric}
\end{equation}
The interval corresponding to this effective metric is obtained from the
inverse metric $g_{\mu\nu}$:
\begin{equation}
ds^2=g_{\mu\nu}dx^\mu dx^\nu=-dt^2\left(1-{v^2\over (c^{x} )^2}\right) -2 dxdt
{v\over (c^{x})^2}+ {dx^2\over (c^{x} )^2}+{dy^2\over c_\perp^2}+ {dz^2\over
v_F^2}~.
 \label{Interval}
\end{equation}
Two event horizons occur at points $x=\pm x_h$, where both $g^{xx}=0$ and
$g_{00}=0$,
\begin{equation}
\tanh  {x_h\over d} =  {v\over c_\perp} ~.
 \label{Horizon}
\end{equation}
At these points the
speed of particles in the
$x$-direction equals the velocity of the soliton in the same directon,
$|c^{x}(\pm x_h)|=  v$, Fig.~1. Between the horizons,
$|c^{x}|$ is less than $v$,  thus particles cannot propagate out across the
horizon at $x=x_h$. This horizon represents the black hole. Another horizon, at
$x=-x_h$, corresponds to the white hole, since the particles cannot propagate
into the region behind horizon.

So, this domain wall represents the nonrotating black hole with the singularity
at $x=0$. Moreover, as distinct from the extended description of the black
hole,
both past and future horizons are physical. In
condensed matter the event horizon appears only if the effective metric
is non-static, i.e.  $g_{0i}\neq 0$. Is this constraint valid for the
Planck condensed matter too?

\subsection{Hawking temperature and Bekenstein entropy}

In the presence of a horizon the vacuum becomes ill-defined. This can be seen
from thermodynamic functions of ``relativistic" fermions
in the moving soliton. For example, let us consider the normal
component of the liquid (``matter"), which  comoves   with the soliton and has
 temperature
$T$. Then the local pressure of this system\cite{Casimir}
\begin{equation}
P=\hbar {7 \pi^2 \over 180} ~{T^4\over  g^2_{00}\hbar^4}
(-g)^{1/2}={7 \pi^2 \over 180 \hbar^3}~  {\tilde T^4\over v_Fc_\perp
c^x}~~,~~\tilde T = T
\left(1-{v^2\over  (c^x)^2 }\right)^{-1/2} ~~,
\label{Pressure}
\end{equation}
diverges at horizon together with the effective temperature $\tilde T$. Thus
there is no equilibrium thermodynamic state in the presence of horizon.   The
existence of a horizon is associated with the dissipative state
characterized by
the Hawking radiation. The latter is thermal black-body radiation, with
temperature determined by the ``surface gravity":
\begin{equation}
T_H={\hbar\over 2\pi k_B} \kappa~,~\kappa= \left({dc^x\over dx}\right)_h
~~.
\label{HawkingT}
\end{equation}
In our case the Hawking temperature depends on the velocity $v$:
\begin{equation}
T_H(v)=T_H(v=0) \left( 1- {v^2\over c_\perp^2}\right)~~~,~~~T_H(v=0)={\hbar
c_\perp
\over 2\pi k_B d}~
\label{T(v)}
\end{equation}
The Hawking radiation can be detected by quasiparticle detectors. Also it
leads to the energy dissipation and thus to deceleration of the moving domain
wall even if the real temperature $T=0$. It would be interesting to find out
whether the dissipation disappears and the velocity of the domain wall is
stabilized if the real temperature of superfluid equals the Hawking
temperature,
$T=T_H(v)$. Some hint on such a possibility can be found in the paper on
the moving boundary between $^3$He-A and  $^3$He-B.\cite{SchopohlWaxman}

Due to deceleration caused by Hawking radiation, the Hawking temperature
increases with time. The distance between horizons, $2x_h$, decreases until the
complete stop of the domain wall when the two horizons merge. In a similar
manner, shrinking of the black hole can stop after the Planck scale is
reached. The  Hawking temperature approaches its asymptotic value
$T_H(v=0)$ in Eq.(\ref{T(v)}), but when the horizons merge, the Hawking
radiation disappears, since the stationary domain wall is nondissipative.
One can show \cite{JacobsonVolovik2}, that the entropy of the domain wall
does not disappear, but approaches a finite value, which
corresponds to  one degree of quasiparticle freedom per Planck area. This is
similar to the  Bekenstein entropy, but it comes from the ``nonrelativistic''
physics at the ``trans-Planckian'' scale. Nonzero entropy results from the
fermion zero modes: bound states at the domain wall with exactly zero energy.
Such bound states, dictated by topology of the texture, are now intensively
studied in high-temperature superconductors and other unconventional
superconductors/superfluids (see references in \cite{Calderon} and
\cite{1/2vortex}). Event horizons with Hawking radiation and Bekenstein entropy
also appear in the core of vortices.\cite{RotatingCore} The influence of the
high-energy nonrelativistic spectrum on Hawking radiation is under
investigation (see \cite{Corley}).

\section{Metric Induced by Quantized Vortices}

The linear and point-like  topological defects also induce effective metrics,
which can be interesting for the theory of gravitation. The simulation of
2D and
3D conical singularities by disgyrations and monopoles in $^3$He-A one can find
in Ref.\cite{GravityOfMonopole}. Here we consider quantized vortices in
superfluid
$^3$He-A films.

\subsection{Vortex as cosmic spinning string}

The vortex in thin films, where the ${\hat {\bf l}}$-vector is fixed, is
a topologically stable object with the superfluid velocity circulating
around the
origin,
${\bf v}_s=N\kappa\hat{\bf \phi}/2\pi r$. Here $\kappa=\pi\hbar/2m_3$ is the
quantum of circulation ($m_3$ is the mass of the $^3$He atom), and $N$ is an
integer or half-odd integer circulation quantum number. This superflow induces
the effective space, where fermions are propagating along geodesic curves, with
the interval\cite{VortexVsSpinningString}
\begin{equation}
  ds^2=-\left(1-{v_s^2(r)\over c_\perp^2}  \right) dt^2 -  {\hbar N \over
 m_3 c_\perp^2}d\phi dt   +{1\over c_\perp^2}(dr^2+
  r^2d\phi^2)+{1\over v_F^2} dz^2
 ~.
\label{IntervalInVortex}
\end{equation}
Far from the vortex axis, where $v_s\ll c_\perp$, this transforms to the metric
induced by the cosmic spinning string. This similarity leads to the
gravitational Aharonov-Bohm effect and as a result to an Iordanskii type
lifting force acting on the vortex (in the presence of quasiparticles) and on
the spinning string (in the presence of matter).\cite{VortexVsSpinningString}

\subsection{Instability of ergoregion}

Another important property of the  metric induced by the vortex is the
existence
of the so-called ergoregion, where quasiparticles have negative energy:
$E({\bf p}) +{\bf p}\cdot{\bf v}_s<0$. The ergoregion occupies
a cylinder of radius $r_e=\hbar N/2m_3c_\perp$, where the flow velocity is
larger than the speed of light, $v_s> c_\perp$, and thus $g_{00}<0$
(Fig.2a). The flow is circulating along the boundary of the
ergoregion, so the horizon is absent here: the ergosurface is thus
separated from
the horizon. A similar situation occurs for the rotating black hole; and it
was argued that  the ergoregion can be unstable
there\cite{QuantumErgoregionInstablity}  Here we provide a condensed matter
illustration of such an instability.   It was shown \cite{CriticalVelocity}
that
for the superfluids with  ``relativistic'' fermions, such as superfluid
$^3$He-A, the superfluid vacuum is unstable if $v_s$ exceeds the
``speed of light''.  In the vortices with small $N$ such an instability is
prevented by the gradients of the order parameter in the core. But if the
vortex
winding number is big,
$N\gg 1$, so that the radius of the ergoregion essentially exceeds the
superfluid
coherence length,
$r_e=\hbar N/2m_3c_\perp \gg \hbar/m_3 c_\perp \sim \xi$, the instability can
develop.

The development of the vacuum instability is also an interesting
phenomenon. In cosmology it can result in the inflationary expansion of the
Universe. In principle we can also construct the situation in which the
``speed of light'' decays exponentially or as a power law faster than $1/t$,
and thus simulate the inflation and investigate the growth of perturbations
during inflation.

In the case of the vortex with $N \gg 1$, the vacuum instability in the
ergoregion results in the splitting of the
$N$-quantum vortex into $N$ singly quantized vortices.  However it is
interesting
to consider what happens if we impose the condition that the axial symmetry of
the vortex is preserved. In this case the resulting equilibrium core structure
is shown in Fig.2b. The thin shell, with the thickness of order the
coherence length (Planck length), which consists  of the normal
(non-superfluid)
state,  separates two superfluid vacua. In the inner vacuum the superflow is
absent; as a result the metric experiences a jump across the shell. The
shell is
situated at the radius
$r_c >r_e$, so that the ergoregion is completely erased: both inner and outer
vacua do not contain ergoregions. Thus an equilibrium vacuum does not sustain
ergoregions.

Another problem is related to
dependence of the vacuum on the reference frame. Due to Galilean invariance,
there are two sources of the same non-static metric, say, in
Eq.(\ref{IntervalInVortex}). One of them is the superfluid velocity
field ${\bf v}_s({\bf r})$; another one appears in the moving texture in the
comoving frame. The vacuum is the same
for  two metrics, unless a  horizon or an  ergoregion appear,   similar to
ambiguity of relativistic quantum vacuum  in the presence
of horizon. This problem includes  properties of rotating quantum vacuum (see
e.g.
\cite{RotatingQuantumVacuum}) and   can be investigated in
$^3$He-A using  rotating cylinders.

\section{Conclusion}

Using superfluid phases of $^3$He one can simulate a broad spectrum of
phenomena related to the quantum vacuum. We discussed only few of them.
On the conceptual level the lessons of $^3$He-A suggest:  The gravity is
the property of quantum ``Planck matter" in the classical low-energy limit.
Gravitational field arise as collective modes of the dynamical deformations of
topologically stable gap nodes. The same topological stability provides the
zero
mass of neutrino, if the overall topological invariant of the ``Planck matter"
is nonzero. The equilibrium state of ``Planck matter" does not gravitate,
suggesting possible route to solution of cosmological constant problem. The
vacuum can be highly unstable behind horizon, and thus can resist to formation
of black holes. Fundamental constants  $G$  and  $c$ should  depend on
temperature; what about the third fundamental constant -- Planck constant
$\hbar$?

\vfill\eject

Figure Captions:

1. Solid line is the ``speed of light" in the $x$-direction, $c^x(x)$, for
the domain wall moving with velocity $v$ in thin film of
$^3$He-A. The speed of light crosses zero at $x=0$. For the moving wall the
black and white hole pair appears for any velocity $v$ below $|c^x(\infty)|$.
At horizons $g^{xx}(\pm x_h)=0$ and $g_{00}(\pm x_h)=0$. Arrows show possible
directions of propagation of particles.

2. (a) Regular structure of the  core of the vortex. In the region
$r<r_e$ the superfluid velocity $v_s$ exceeds the speed of
``light'', $c_\perp$, and $g_{00}>0$. In this ergoregion the quasiparticle
energy
is negative. For vortices with large winding number $N\gg 1$, the quantum
superfluid vacuum is unstable in the ergoregion. (b) For the axially symmetric
vortex this instability leads to the reconstruction of the core. In the new
core
structure, the normal shell separates two superfluid vacua, in which
superfluid velocity does not exceed the speed of
``light''. Ergoregion is absent.

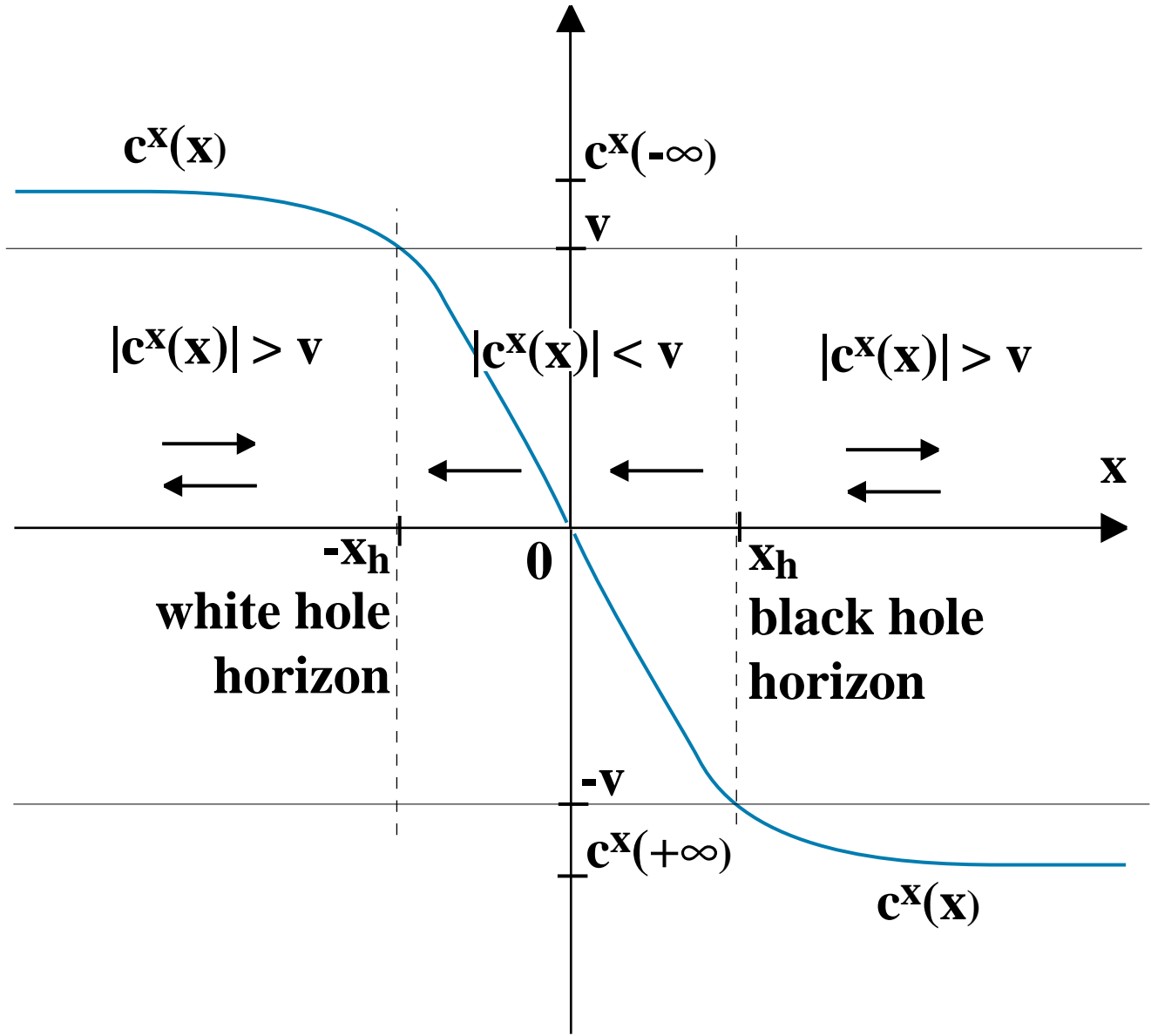
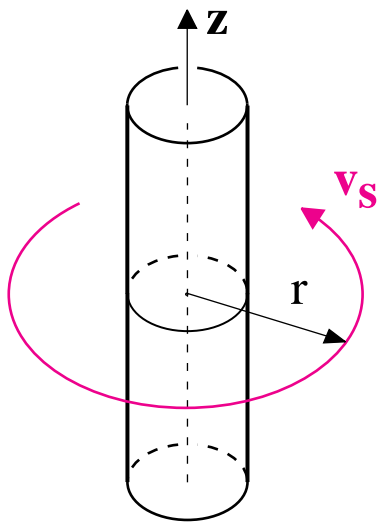
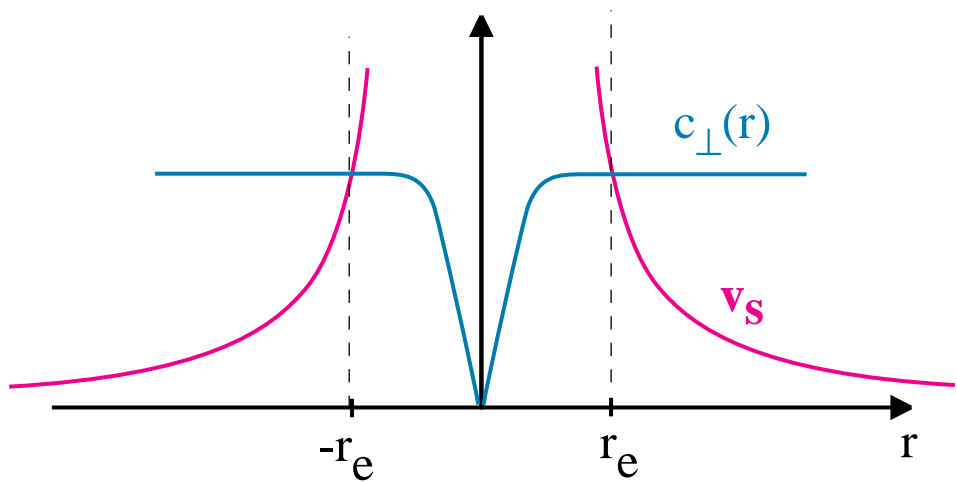
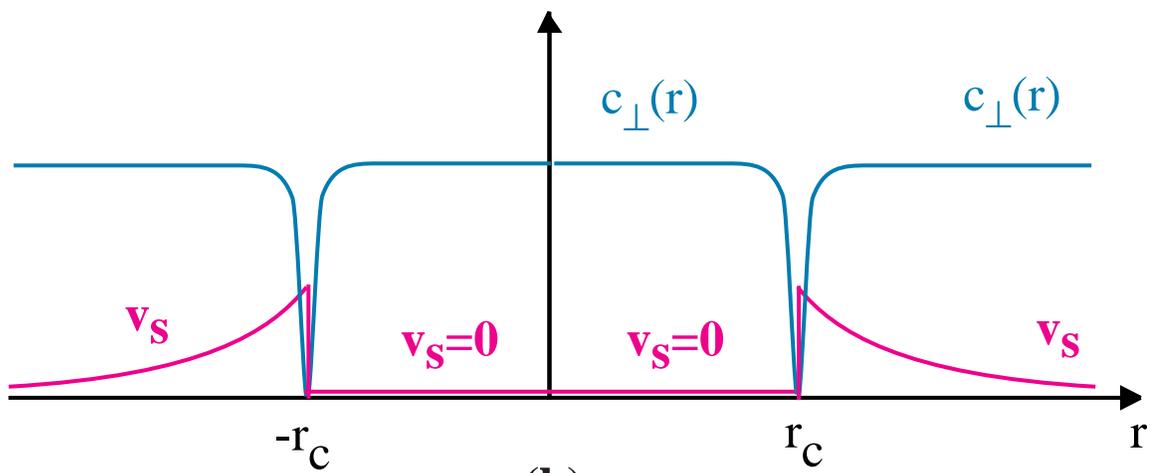
\end{document}